\def\tipo{1}

\ifnum \tipo = 1
 \documentstyle[pre,twocolumn,aps,psfig]{revtex}
 \def\figsize{8.5cm}
 \def\figsiz1{8.5cm}
 \def \frontmatter{\twocolumn[\hsize\textwidth\columnwidth\hsize\csname@twocolumnfalse\endcsname}
 
\else
 \documentstyle[preprint,aps,pre,psfig]{revtex}
 \def\figsize{12cm}
 \def\figsiz1{12cm}
 \def\frontmatter{}
 
\fi

\begin{document}
\draft
\frontmatter
\title{Relation between Magnitude Series Correlations and
 Multifractal Spectrum Width} 
\author{Yosef~Ashkenazy$^{\text{1}}$, Shlomo~Havlin$^{\text{2}}$, Plamen~Ch.~Ivanov$^{\text{1,3}}$, \\
 Chung-K. Peng$^{\text{3}}$, Verena Schulte-Frohlinde$^{\text{1}}$,
and H. Eugene Stanley$^{\text{1}}$}
\address{ 
$^1$ Center for Polymer Studies and Department of Physics, Boston
University, Boston, Massachusetts 02215, USA\\
$^2$ Gonda-Goldschmied Center and Dept. of Physics, Bar-Ilan University, 
Ramat-Gan, Israel
$^3$ Beth Israel Deaconess Medical Center,
Harvard Medical School, Boston,\\Massachusetts 02215, USA\\
}
\date{\today}
\maketitle
\begin{abstract}
{ 
We study the correlation properties of long-range correlated time
series, $x_i$, with tunable correlation exponent and built-in
multifractal properties.  
For the cases we investigate we find that the correlation exponent of the
magnitude series, $|x_{i+1}-x_i|$, is a monotonically increasing
function of the multifractal spectrum width of the original series. 
}
\end{abstract}
\ifnum \tipo = 1
]
\fi

\def\figureI{
\begin{figure}[thb]
\centerline{\psfig{figure=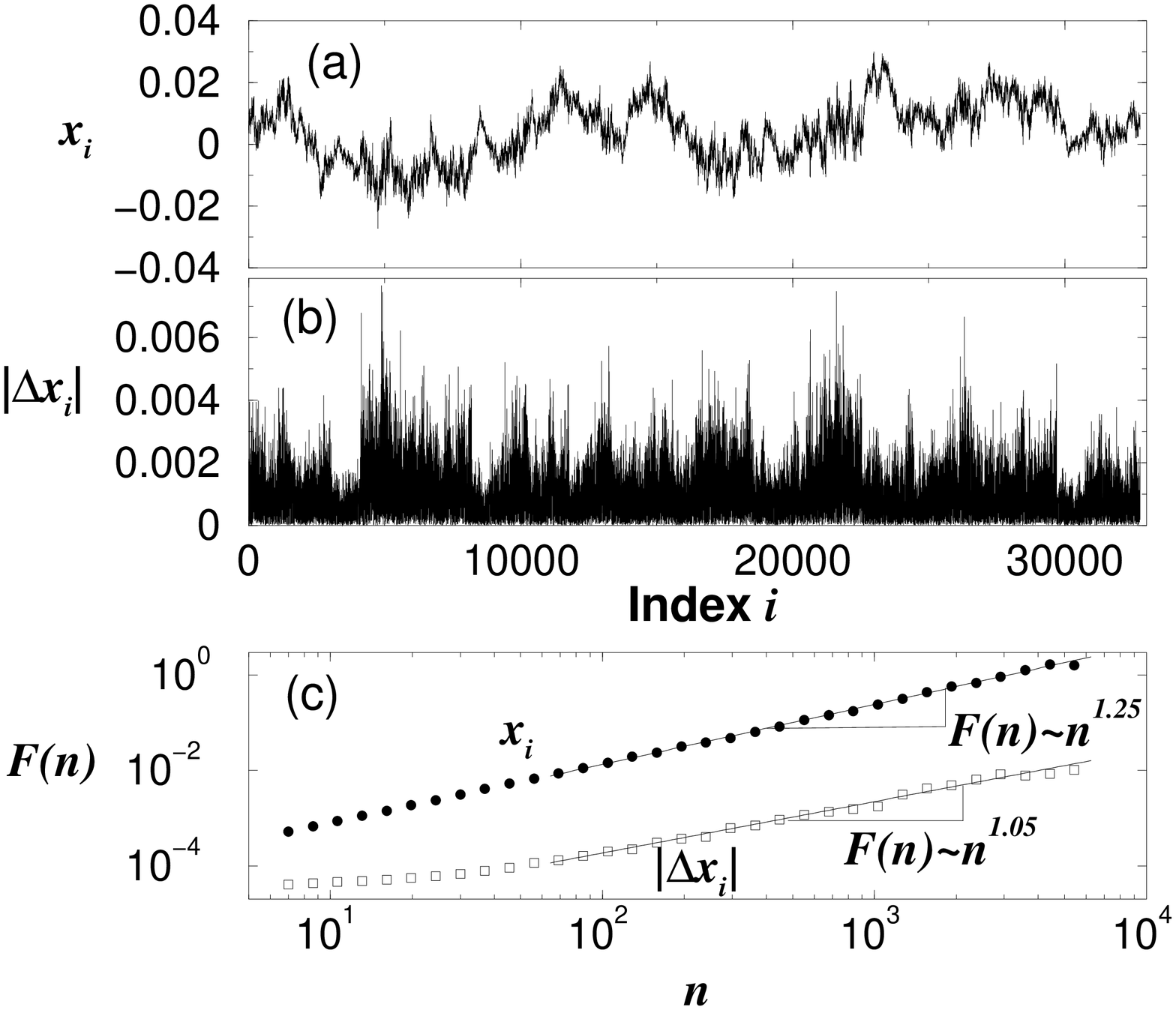,width=\figsiz1,angle=-90}}
{
\ifnum\tipo=2
\vspace*{0.0truecm}
\fi
\caption{\label{fig1} 
(a) An example of 32,768 data points of an artificial MF random series 
with Hurst exponent of $H=0.25$. We use the log-normal $W$ wavelet 
cascade with $\mu=-(\ln 2)/4$ and $\sigma=0.1$ to generate the artificial 
series.
(b) The magnitudes of increments obtained from the original series shown in (a).
Patches of more ``volatile'' increments with large magnitude are
followed by patches of less volatile increments 
with small magnitude. In contrast, a monofractal series in (a) would
result in a homogeneous series for the magnitudes of increments 
$|\Delta x_i|$.
(c) The root mean square fluctuations $F(n)$ versus the window size $n$
(obtained by the second order detrended fluctuation analysis 
\protect\cite{Peng94}) for the 
original series shown in (a) and for the magnitude series shown in
(b). The scaling exponents are calculated for window scales
$n>64$. The correlations of the original series follow a scaling law
$F(n) \sim n^{1.25}$ and is consistent with a theoretical Hurst exponent
of $H=0.25$ since $H=\alpha-1$ (the difference of 1 is due to integration 
in the DFA procedure). 
The magnitude series is also correlated with an exponent 
of $\alpha \approx 1$. 
}}
\end{figure}
}

\def\figureII{
\begin{figure}[thb]
\centerline{\psfig{figure=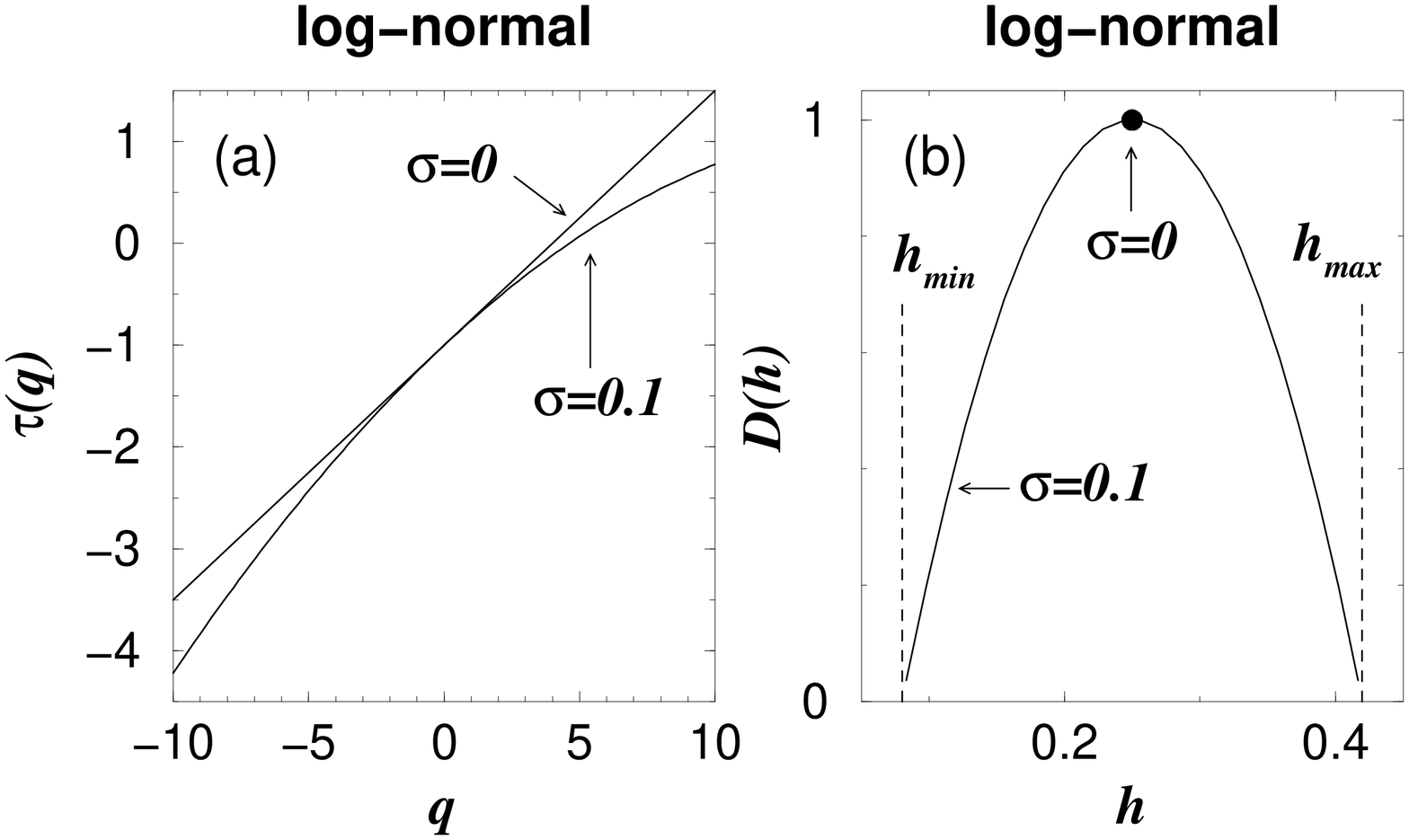,width=\figsiz1,angle=-90}}
\centerline{\psfig{figure=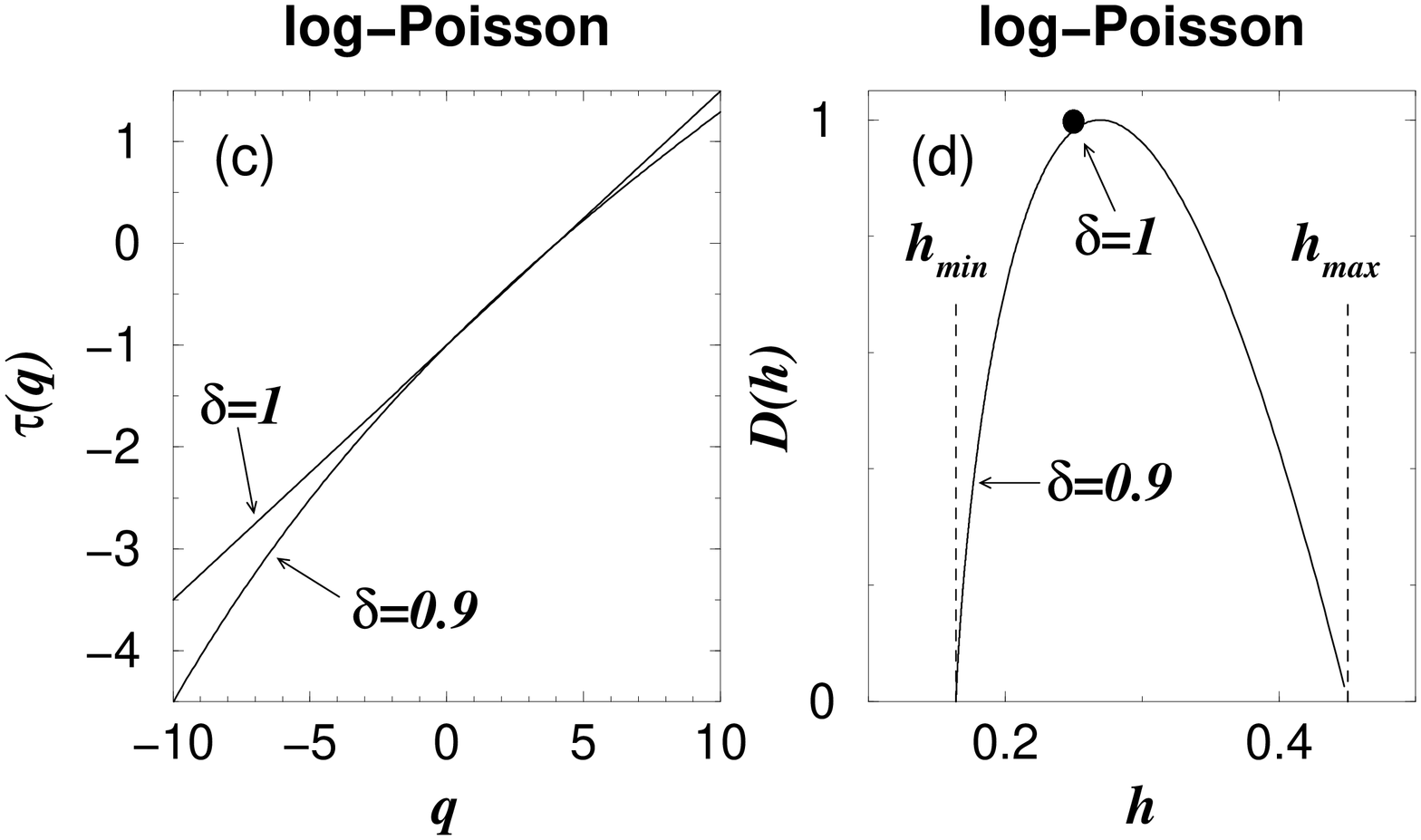,width=\figsiz1,angle=-90}}
{
\ifnum\tipo=2
\vspace*{0.0truecm}
\fi
\caption{\label{fig2} 
(a) Theoretical scaling exponent $\tau (q)$ for $\sigma=0$ and 
$\sigma=0.1$ for a MF series generated using random variable $W$
from a log-normal distribution. The series has symmetric MF spectra
where the width of the MF spectrum is proportional to $\sigma$ [Eq. 
(\protect\ref{e2})].  For $\sigma=0$ $\tau(q)$ is linear and the series
is monofractal while for $\sigma=0.1$ $\tau(q)$ is curved and the series 
is MF.
(b) The MF spectrum, $D(h)$, obtained from $\tau(q)$ shown in (a). For 
$\sigma=0$ ($\bullet$) the $D(h)$ MF spectrum is a single point indicating
monofractal behavior with a global Hurst exponent $H=0.25$. For 
$\sigma=0.1$ (curved solid line) the $D(h)$ spectrum
is wide indicating MF behavior. The MF spectrum is symmetric and
centered around $h=0.25$. The vertical dashed lines indicate the minimum
($h_{min}$) and maximum ($h_{max}$) values of the local Hurst exponent.
(c) Same as (a) for a MF series generated using random variable $W$ 
from a log-Poisson distribution. In this case the MF spectrum
is {\it asymmetric} for which the second moment $\tau(2)$ and the forth moment 
$\tau(4)$ are almost fixed. The $\tau(q)$ spectrum is given for $\delta=1$
(linear dependence that indicates monofractality) and for $\delta=0.9$
(curved line that indicates multifractality). Note that $\tau(q)$ remains 
almost unchanged for positive $q$'s while $\tau(q)$ changes
significantly for negative $q$'s.
(d) The MF spectrum, $D(h)$, for the examples shown in (c). The
$D(h)$ spectrum for $\delta=1$ is monofractal ($\bullet$); the global Hurst
exponent is $H=0.25$. For $\delta=0.9$
(solid curved line) $D(h)$ MF spectrum is broad and asymmetric.
$D(h)$ is more stretched to the right indicating larger changes in
the negative moments. 
}}
\end{figure}
}

\def\figureIII{
\begin{figure}[thb]
\centerline{\psfig{figure=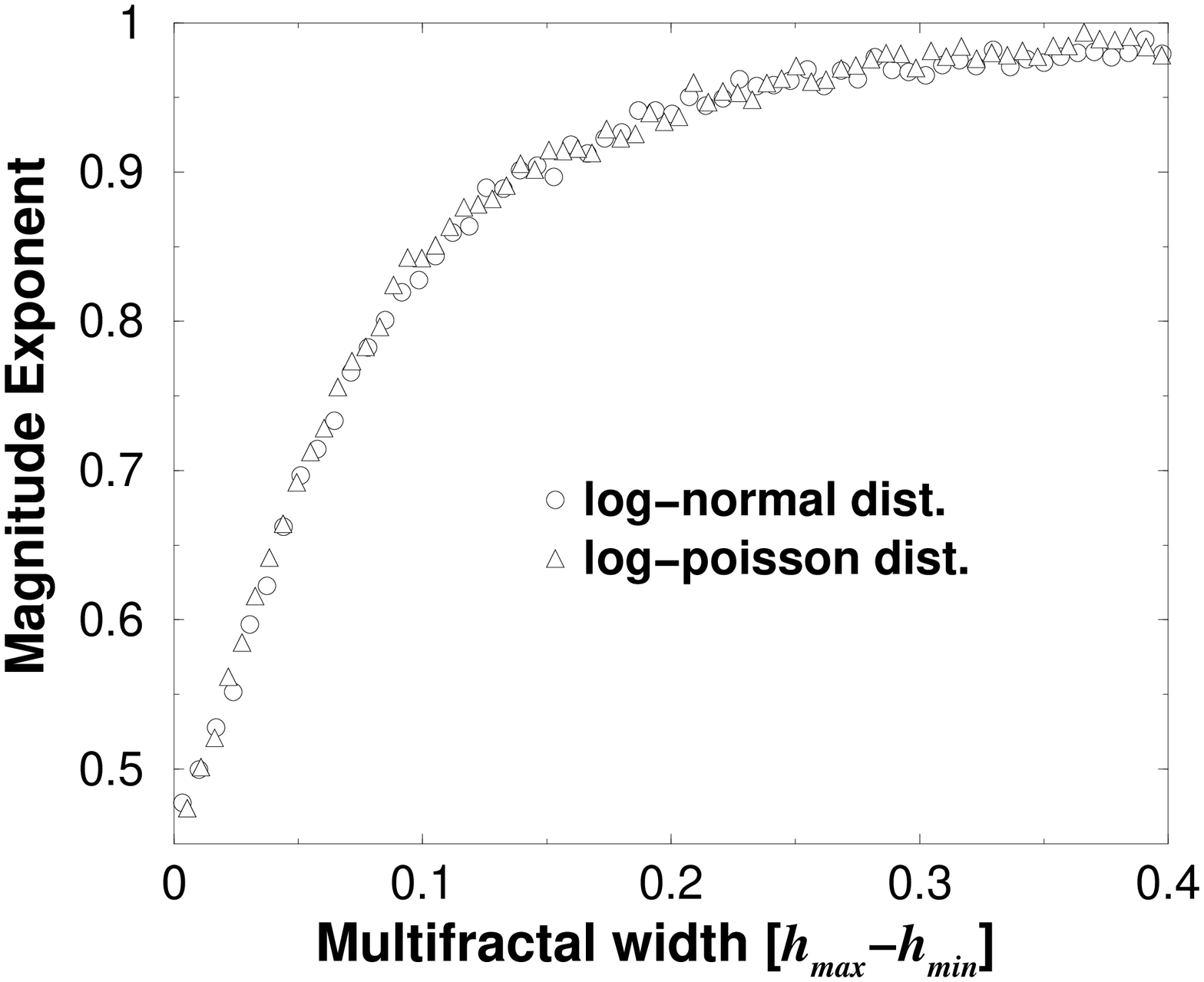,width=\figsize,angle=-90}}
\caption{\label{fig3}
The magnitude scaling exponent versus the analytical MF spectrum width,
$\Delta h \equiv h_{max}-h_{min}$. 
For each point we generate 20 realizations each of 65,536 points. We
calculate the magnitude scaling exponent in the range $64\le n \le
4096$ using the second order DFA; the average exponent is shown (the
standard deviation is less 
than 0.05). Both examples, log-normal distribution ($\circ$) and
log-Poisson distribution ($\triangle$), 
exhibit the same behavior of monotonically increasing magnitude 
exponent as a function of the MF spectrum width. We approximate this 
increase by $1/(1+e^{-17\Delta h})$. 
In contrast, the value of the 2-point scaling exponent of the original
series does not depend on the width of the MF spectrum.
}
\end{figure}
}

Many natural systems exhibit complex dynamics 
characterized by 
long-range power-law 2-point
correlations \cite{Shlesinger,Vicsek,Takayasu}. In some cases the output of 
such systems --- a series of a fluctuating variable $x_i$ --- may be
characterized by a single 2-point correlation exponent, while in
other cases
the series may be quantified by a spectrum of 
exponents called the multifractal (MF) spectrum \cite{Feder,Gene,Arneodo94}. 
The MF spectrum quantifies in details the long-range correlation
properties of a series $x_i$.
Nonetheless, additional way to measure the correlation properties of a
complex series, $x_i$, 
is to analyze the long-range power-law correlations for the magnitudes
of the increments $|\Delta x_i| \equiv |x_{i+1}-x_i|$. 

For example, complex fluctuations in financial indices 
are characterized by a broad MF spectrum
\cite{Mandelbrot}. Moreover, recent reports showed that there are positive 
long-range power-law correlations  
in the volatility (magnitude of the increments) in such financial time series
\cite{Liu}. Biomedical time series are also found to exhibit similar complex 
behavior --- e.g., the human heartbeat dynamics generates fluctuations which 
are anti-correlated on a wide range of time scales \cite{Kobayashi} and
are also characterized by a broad MF spectrum
\cite{Plamen}. Furthermore, it was recently reported that the  
magnitudes of these fluctuations are positively correlated and exhibit 
power-law scaling behavior over scales from seconds to hours
\cite{Ashkenazy2001}. These and  
other examples \cite{GaitAshkenazy,Jan2002} raise the possibility that
there may be a relation  
between the scaling in the magnitude series and the MF spectrum of the 
original series which is independent of the type of long-range correlations 
present in the original series.

Here we study the relation between the scaling exponent which quantifies 
the power-law correlations in the magnitude series and the MF spectrum  
of artificially generated time series with analytically built-in MF 
properties. We find that the scaling exponent of the magnitude series
is a monotonically increasing function of the MF spectrum width of the
original series. 

Long-range power-law correlated series are {\it monofractal} when 
they are characterized by exponents $\tau(q)$ for different moments $q$ being
{\it linearly} depend on $q$,
\begin{equation}
\tau(q)=Hq+\tau(0),
\end{equation} 
with a single Hurst exponent \cite{Hurst}, 
\begin{equation}
H \equiv d\tau/dq = {\rm const}.
\end{equation}
Other long-range correlated series are {\it multifractal} and are 
described by $\tau(q)$ that depend  {\it nonlinearly} on $q$,
\begin{equation}
h(q) \equiv d\tau/dq  \ne {\rm const},
\end{equation}
where $h(q)$ is the local Hurst exponent \cite{Arneodo94}.
The MF spectrum of the series is defined by
\begin{equation}
D(h)=h(q)q-\tau(q),
\end{equation}
where for monofractal series it collapses to a single point,
\begin{equation}
D(H)=-\tau(0).
\end{equation} 
Here we show that the magnitude series scaling exponent is related to
the width of the MF spectrum $D(h)$ of the original series,
\begin{equation}
\Delta h \equiv h_{max}-h_{min},
\end{equation}
where $h_{min}$ and $h_{max}$ are the minimum and maximum of the local 
Hurst exponent.

We use the algorithm proposed in \cite{Arneodo98} to generate
artificial MF series. 
Briefly, the random MF series is built by specifying discrete
wavelet coefficients 
which are constructed recursively using a 
random variable $W$ (see \cite{Arneodo98,Daubechies} for details).
Once the wavelet coefficients are obtained, we apply inverse 
wavelet transform to generate the MF random series; we use the
10-tap Daubechies discrete wavelet transform \cite{Daubechies,NR}. The
MF properties of the obtained artificial series depend on 
the form of distribution from which the random variable $W$ is chosen.

We consider two different types of probability
distributions for the random variable $W$, the log-normal distribution
and the log-Poisson distribution. For these two examples the MF
properties are known analytically \cite{Arneodo98}.
The MF spectrum $D(h)$ is symmetric for the log-normal distribution and
asymmetric for the log-Poisson distribution. 

First we consider the case of a log-normal random variable $W$
(Fig. \ref{fig1}a) --- i.e., $\ln |W|$ 
is normally distributed and $\mu$, $\sigma^2$ are the
mean and the variance of $\ln |W|$.
We choose $\mu=-(\ln 2)/4$. In this case the scaling
exponents, $\tau(q)$, are given by \cite{Arneodo98}
\begin{equation}
\tau(q)=-\frac{\sigma^2}{2\ln 2} q^2 +\frac{q}{4}-1,
\end{equation}
and the MF spectrum $D(h)$ is
\begin{equation} \label{e2a}
D(h)=-\frac{(h-1/4)^2\ln 2}{2\sigma^2}+1.
\end{equation}
We calculate the MF spectrum width by solving
$D(h)=0$ and find
\begin{equation} \label{e2}
\Delta h =\Big(\frac{\sqrt{2}\sigma}
{\sqrt{\ln 2}}+{1 \over 4}\Big) - \Big(-\frac{\sqrt{2}\sigma}
{\sqrt{\ln 2}}+{1 \over 4}\Big) = \frac{2\sqrt{2}\sigma}{\sqrt{\ln 2}}. 
\end{equation}
For $\sigma= 0$, the series is monofractal, while for large value of
$\sigma$ it has a broad MF spectrum. 
The MF spectrum is symmetric around $D(1/4)=1$ and collapses to a single
point $D(1/4)=1$ when $\sigma\to 0$ [Eq. (\ref{e2a}) and 
Fig.~\ref{fig2}a,b].

We generate series with different $\sigma$ values ranging from 0 to
0.1 (20 realizations for each $\sigma$) and calculate the magnitude
scaling exponent using the method of detrended fluctuation analysis (DFA)
\cite{Peng94} (see Fig. \ref{fig1}c and Fig.~\ref{fig3}). We find that
the scaling exponent of the magnitude series increases 
monotonically with $\sigma$ from uncorrelated magnitude series with
scaling exponent $\approx 0.5$ for $\sigma=0$ to strongly correlated 
magnitude series with exponent $\approx 1$ for $\sigma=0.1$. Further, 
the magnitude exponent converges to $1$. 

Next we show that the scaling exponent of the magnitude does {\it
not} only depend on the positive moments (and especially the fourth 
moment) but rather relates to the entire
MF spectrum. For this purpose we study an example with an
asymmetric MF spectrum where the exponents for negative
moments, $h(q<0)$, are changed drastically when the MF
spectrum width is changed, while the exponents for the positive moments
$h(q>0)$ are less significantly changed.
We tune the parameters in such a way that the fourth moment $\tau(4)$ is
fixed and the second moment $\tau(2)$ is hardly changed. 

To generate a series with an asymmetric MF spectrum
we consider the case of a random variable $W$ from a log-Poisson
distribution, where
$\lambda$ is the mean and the variance of a Poisson distributed
variable $P$, and $\ln 
|W|$ has the same distribution function as $P\ln \delta+\gamma$.
We choose $\lambda =\ln 2$ 
and $\gamma = -\delta^4 (\ln 2) /4$ where $\delta$ is an appropriately 
chosen positive parameter \cite{Arneodo98}. Under this choice of parameters,
\begin{equation}
\tau(q) = -\delta^q+q \delta^4/4,
\end{equation}
and
\begin{equation}
D(h)=\frac{1}{\ln \delta} \big( h - \frac{\delta^4}{4}
\big) \Big[\ln\Big( \frac{
h - \delta^4/4}{-\ln \delta }\Big) -1 \Big].
\end{equation}
Thus the fourth moment $\tau(4)$ is independent of $\delta$
\begin{equation}
\tau(4)=0,
\end{equation}
and the width of the MF spectrum depends on $\delta$ 
\begin{equation} \label{e3}
\Delta h =\big(\delta^4/4-e\ln
\delta\big)-\big(\delta^4/4\big)=-e\ln \delta. 
\end{equation}
For $\delta=1$ the MF spectrum collapses to a single
point (monofractal) with Hurst exponent $H=1/4$ and $D(1/4)=1$.
For both cases, decreasing or increasing $\delta$, the MF spectrum becomes
broader. Here we change the MF spectrum width by decreasing
$\delta$. The major change in the MF spectrum occurs for negative moments
$q<0$ (Fig.~\ref{fig2} c, and d). We find that the scaling exponent of 
the magnitude series increases monotonically with the MF spectrum width
(Fig.~\ref{fig3}) although the positive moments are hardly
changed. Thus, the magnitude series scaling exponent is related to the
entire MF spectrum of the original series and not just to the 
positive moments.

We summarize the results of the log-normal and the log-Poisson examples in
Fig.~\ref{fig3} --- the magnitude scaling exponent is plotted versus the
width of the MF spectrum
[Eqs. (\ref{e2}),(\ref{e3})]. Surprisingly, these two  
examples collapse on the same curve. This collapse suggests a possible
one-to-one relation between the magnitude scaling
exponent and the MF spectrum width \cite{remark5}.

The relation between the scaling exponent of the magnitude 
series and the MF spectrum width of the original series is consistent 
with recent works on MF random walk models with built-in correlations
\cite{Bacry01}.
Recently \cite{Ashkenazy2001} it was found that power-law correlations
in the magnitude series are related to nonlinear
\cite{remark-nonlinearity1} properties of the increment series $\Delta
x_i$ --- nonlinear series have correlated magnitude series while
linear series have uncorrelated magnitude series. On the other hand,
monofractal series are linear series while MF series series are
nonlinear \cite{Plamen}. Thus the origin of the magnitude series
correlations and the corresponding MF spectrum may be related through
the nonlinearity of the original time series.

\def\out{
Studies of econometric time series and human heartbeat interval series
have treated and analyzed separately the magnitude (volatility)
correlations and the MF spectrum. Moreover, models were developed to
satisfy one property without considering the other.  The link between
the magnitude series correlations and the MF spectrum
suggest that models (or complex systems) with long-range correlations
in the magnitude (volatility) series likely implies a broad MF
spectrum while absence of correlations in the magnitude series likely
implies a monofractal spectrum \cite{GaitAshkenazy,remark5}. 
}

The results described here may be important from a practical
point of view. The calculation of the MF spectrum from a
time series involves advanced numerical techniques
\cite{Arneodo94} and requires long time series. Our analysis
of the magnitude series is less sophisticated than the MF analysis and
is applicable to shorter time series than one need for the MF
analysis. We note that a direct numerical calculation of the MF 
spectrum width from a time series may not follow the monotonic relation
described in Fig.~\ref{fig3} due to overestimation of the
MF spectrum width caused by (i) the numerical technique for
calculating the MF spectrum [and estimation of $\tau(q\to \pm
\infty)$] and by (ii) finite series length. However, we expect the
magnitude scaling exponent to increase monotonically with the width of the
MF spectrum \cite{GaitAshkenazy,remark5} since both are related to the
nonlinearity of the original series. 

Partial support was provided by the NIH/National Center for Research
Resources (P41 RR13622), the Israel-USA 
Binational Science Foundation, and the German Academic Exchange Service
(DAAD). We thank A.L. Goldberger and J.W. Kantelhardt for helpful
discussions.

\ifnum\tipo=1
  \figureI
  \figureII
  \figureIII
\fi
\ifnum\tipo=2
  \figureI
  \figureII
  \figureIII
\fi


\begin{references}

\bibitem{Shlesinger} M.F. Shlesinger, {Ann. NY Acad. Sci.} {\bf 504},
214 (1987). 

\bibitem{Vicsek} T. Vicsek, {\it Fractal Growth Phenomenon} 2nd edn.
(World Scientific, Singapure, 1993).

\bibitem{Takayasu} H. Takayasu, {\it Fractals in the Physical Sciences}
(Manchester Univ. Press, Manchester, UK, 1997).

\bibitem{Feder} J. Feder, {\it Fractals} (Plenum Press, New York, 1988).

\bibitem{Gene} H.E. Stanley, Nature {\bf 405}, 335 (1988).

\bibitem{Arneodo94} J.F. Muzy, E. Bacry, and A. Arneodo,
Int. J. Bifurcat. Chaos {\bf 4}, 245 (1994).

\bibitem{Mandelbrot} B.B. Mandelbrot, {\it Fractals and Scaling in
Finance} (Springer, New York, 1997). 

\bibitem{Liu} Y.H. Liu, P. Gopikrishnan, P. Cizeau, M. Meyer,
C.-K. Peng, and H.E. Stanley, Phys. Rev. E {\bf 60}, 1390 (1999).

\bibitem{Kobayashi} M. Kobayashi and T. Musha, { 
IEEE Trans. Biomed. Eng.} {\bf 29}, 456 (1982).

\bibitem{Plamen}
P.Ch. Ivanov, {\it et al.}, Nature {\bf 399}, 461 (1999).

\bibitem{Ashkenazy2001} Y. Ashkenazy, P.Ch. Ivanov, S. Havlin,
C.-K. Peng, A.L Goldberger, and H.E. Stanley, Phys. Rev. Lett. {\bf
86}, 1900 (2001). 

\bibitem{GaitAshkenazy} Y. Ashkenazy, J.M. Hausdorff, P.Ch. Ivanov,
and H.E. Stanley, preprint (cond-mat/0103119).

\bibitem{Jan2002}
J.W. Kantelhardt, {\it et al.}, Phys. Rev. E {\bf 65}, 051908 (2002).

\bibitem{Hurst} H.E. Hurst, Transactions of 
the American Society of Civil Engineering {\bf 116}, 770 (1951).

\bibitem{Arneodo98} A. Arneodo, E. Bacry, and J.F. Muzy, J. Math. Phys. 
{\bf 39}, 4142 (1998); A. Arneodo, S. Manneville, and J.F. Muzy, Euro. 
Phys. J. B {\bf 1}, 129 (1998).

\bibitem{Daubechies} I. Daubechies, {\it Ten Lectures on Wavelets} (SIAM, 
Philadelphia, PA, 1992). 

\bibitem{NR} W.H. Press, S.A. Teukolsky, W.T. Vetterling, and B.P. 
Flannery, {\it Numerical Recipes in C} 2nd edn., (Cambridge University 
Press, Cambridge, 1995).

\bibitem{Peng94} C.-K. Peng, S.V. Buldyrev, S. Havlin, M. Simons, H.E. 
Stanley, and A.L. Goldberger, Phys. Rev. E {\bf 49}, 1685 (1994).

\bibitem{remark5} We also performed the same procedure (as in Figs. 
\protect\ref{fig1},\protect\ref{fig3}) for a deterministic multifractal 
object, namely, the generalized devil's staircase
\protect\cite{Arneodo94}. 
Also here we find that magnitude exponent  increases monotonically with 
the multifractal spectrum width. However, for very small multifractal 
width the magnitude exponent was $\sim 0$ and then increased sharply to 
the behavior shown in Fig.~\protect\ref{fig3}; this behavior might be 
related to the deterministic nature of this multifractal object. We also 
note that counter examples for which broad MF spectrum is not associated 
with magnitude series correlations (or vice versa) may exist.

\bibitem{Bacry01} E. Bacry, J. Delour, and J.F. Muzy, Phys. Rev. E {\bf 
64},026103 (2001). 

\bibitem{remark-nonlinearity1}  We 
define a process to be {\it linear}
if it is possible to reproduce its statistical properties (such as the
third moment) from the power spectrum and the probability distribution
alone, regardless of the Fourier phases \protect\cite{Schreiber00}.  This
definition includes autoregression processes
($x_{n}=\sum_{i=1}^Ma_ix_{n-i}+\sum_{i=0}^Lb_i\eta_{n-i}$ where $\eta$
is Gaussian white noise) and fractional Brownian motion; the output,
$x_n$, of these processes may undergo monotonic nonlinear
transformations $s_n=s(x_n)$ and still be linear. Processes which do 
not follow the above rule are defined as {\it nonlinear}. 

\bibitem{Schreiber00} T. Schreiber and A. Schmitz, Physica D {\bf 142},
346 (2000).

\end{references}
\end{document}